\documentclass[sigconf,nonacm]{acmart}
\AtBeginDocument{%
  }

\usepackage{titlesec}

\definecolor{mygreen}{RGB}{0,150,0}
\definecolor{myred}{RGB}{255,0,0}
\usepackage{enumitem}

\usepackage[font=small,labelfont=bf]{caption}
\usepackage{array,ragged2e}
\newcolumntype{P}[1]{>{\RaggedRight\arraybackslash}p{#1}}

\definecolor{paired-light-blue}{RGB}{198, 219, 239}
\definecolor{paired-dark-blue}{RGB}{49, 130, 188}
\definecolor{paired-light-orange}{RGB}{251, 208, 162}
\definecolor{paired-dark-orange}{RGB}{230, 85, 12}
\definecolor{paired-light-green}{RGB}{199, 233, 193}
\definecolor{paired-dark-green}{RGB}{49, 163, 83}
\definecolor{paired-light-purple}{RGB}{218, 218, 235}
\definecolor{paired-dark-purple}{RGB}{117, 107, 176}
\definecolor{paired-light-gray}{RGB}{217, 217, 217}
\definecolor{paired-dark-gray}{RGB}{99, 99, 99}
\definecolor{paired-light-pink}{RGB}{222, 158, 214}
\definecolor{paired-dark-pink}{RGB}{123, 65, 115}
\definecolor{paired-light-red}{RGB}{231, 150, 156}
\definecolor{paired-dark-red}{RGB}{131, 60, 56}
\definecolor{paired-light-yellow}{RGB}{231, 204, 149}
\definecolor{paired-dark-yellow}{RGB}{141, 109, 49}

\definecolor{bg1}{HTML}{FF9966}
\definecolor{bg2}{HTML}{CCE5FF}
\definecolor{bg3}{HTML}{FFCC99}
\definecolor{bg4}{HTML}{FFC107}
\definecolor{bg5}{HTML}{FFCCCC}
\definecolor{bg6}{HTML}{D5E8D4}
\definecolor{bg7}{HTML}{eeeeee}
\definecolor{bg8}{HTML}{cdeb8b}
\definecolor{bg9}{HTML}{dae8fc}
\definecolor{bg10}{HTML}{a2e6eb}

\definecolor{bg31}{HTML}{FFCDD2} % light pink

\definecolor{bg32}{HTML}{F8BBD0}

\definecolor{bg33}{HTML}{E1BEE7} % lavender

\definecolor{bg34}{HTML}{D7CCC8} % light tan

\definecolor{bg35}{HTML}{B2DFDB} % light teal

\definecolor{bg36}{HTML}{A5D6A7} % light green

\definecolor{bg37}{HTML}{FFF9C4} %light yellow

\definecolor{bg38}{HTML}{FFECB3} % peach

\definecolor{bg111}{HTML}{CB6843}

\definecolor{bg112}{HTML}{D77C5C}

\definecolor{bg113}{HTML}{E28E6E}
\definecolor{bg114}{HTML}{E89F7D}
\definecolor{bg115}{HTML}{EDAE8A}
\definecolor{bg116}{HTML}{F0BA95}
\definecolor{bg117}{HTML}{F3C29F}
\definecolor{bg118}{HTML}{F6CCAA}
\definecolor{bg119}{HTML}{F8D5B3}
\definecolor{bg120}{HTML}{FADCBD}
\definecolor{bg121}{HTML}{FCE6C7}

\definecolor{bg39}{HTML}{FFE0B2} % apricot

\definecolor{bg40}{HTML}{3CB371} % blush pink

\definecolor{bg43}{HTML}{ffe5d9}

\definecolor{bg15}{HTML}{7FFFD4}

\definecolor{bg17}{HTML}{F0FFFF}

\definecolor{bg18}{HTML}{F5FFFA}

\definecolor{bg19}{HTML}{F8F8FF}

\definecolor{bg20}{HTML}{FFFFFF}

\definecolor{bg21}{HTML}{E1F5FE}

\definecolor{bg22}{HTML}{B3E5FC}

\definecolor{bg23}{HTML}{81D4FA}

\definecolor{bg24}{HTML}{4FC3F7}

\definecolor{bg25}{HTML}{29B6F6}

\definecolor{bg26}{HTML}{03A9F4}
\definecolor{bg27}{HTML}{039BE5}
\definecolor{bg28}{HTML}{0288D1}
\definecolor{bg29}{HTML}{0277BD}
\definecolor{bg30}{HTML}{01579B}
\definecolor{bg16}{HTML}{FFCC99} 
\definecolor{pg51}{HTML}{E8F5E9} % pale green
\definecolor{pg52}{HTML}{C8E6C9} % honeydew green
\definecolor{pg53}{HTML}{B9F6CA} % light mint green
\definecolor{pg54}{HTML}{A9DFBF} % pale sage green
\definecolor{pg55}{HTML}{BCF5A6} % lemon green

\definecolor{pg56}{HTML}{BEF1CE} % seashell green
\definecolor{pg57}{HTML}{CEF6EC} % icy green
\definecolor{pg58}{HTML}{B7F0B1} % feijoa green
\definecolor{pg59}{HTML}{B1F2B5} % pastel light green
\definecolor{pg60}{HTML}{9DF3C4} % greenish cyan

\definecolor{pg61}{HTML}{DEF7E0} % pale green
\definecolor{pg62}{HTML}{E8F8DC} % greenish beige

\definecolor{pg63}{HTML}{EBF7E7} % seafoam green
\definecolor{pg64}{HTML}{F0FDF4} % pale turquoise

\definecolor{pg65}{HTML}{F1FEE7} % mint cream
\definecolor{pg66}{HTML}{F7FFF6} % foam green
\definecolor{pg67}{HTML}{FCFFE7} % pale spring bud
\definecolor{pg68}{HTML}{F4FFD2} % light lime green
\definecolor{pg69}{HTML}{EEFFE2} % tea green
\definecolor{pg70}{HTML}{E3FDF5} % tropical green

\definecolor{connect-color}{RGB}{0,0,0}
\definecolor{middle-color}{RGB}{255,255,255}
\definecolor{leaf-color}{RGB}{173,216,230}
\definecolor{line-color}{RGB}{25,25,112}

\usepackage{url}            % simple URL typesetting
\usepackage{booktabs}       % professional-quality tables
\usepackage{amsfonts}       % blackboard math symbols
\usepackage{nicefrac}       % compact symbols for 1/2, etc.
\usepackage{microtype}      % microtypography
\usepackage{amsmath}
\usepackage{caption}

\usepackage{multirow}
\usepackage{subcaption}
\usepackage{listings}
\usepackage{tcolorbox}
\usepackage{xspace}
\usepackage{makecell}
\usepackage{soul}               % to strikethrough
\setstcolor{red}            %to strikethrough in red

\usepackage{longtable}
\usepackage{tablefootnote}

\usepackage[edges]{forest}
\definecolor{hidden-draw}{RGB}{20,68,106}
\definecolor{hidden-pink}{RGB}{255,245,247}
\definecolor{red}{RGB}{255,0,0}

\definecolor{hidden-draw}{RGB}{0,0,0}
\definecolor{hidden-pink}{RGB}{255,182,193}

\usepackage[T1]{fontenc}

\usepackage[utf8]{inputenc}

\usepackage[draft,textsize=footnotesize,textwidth=15mm]{todonotes}

\usepackage{forest}    % Required for the forest environment

\usepackage{tikz}      % Required for styling nodes

\usepackage{hyperref}  % Required for hyperlinks

\tikzset{
    root style/.style={
        draw,
        rounded corners,
        fill=blue!30, % Color for the root node
        align=center,
        font=\bfseries
    },
    child style/.style={
        draw,
        rounded corners,
        fill=green!30, % Color for child nodes
        align=center,
        font=\bfseries
    },
    grandchild style/.style={
        draw,
        rounded corners,
        fill=red!30, % Color for grandchild nodes
        align=center,
        font=\bfseries
    }
}

\tikzset{
  my-box/.style={
    rectangle,
    draw=hidden-draw,
    rounded corners,
    text opacity=1,
    minimum height=1.5em,
    minimum width=40em,
    inner sep=2pt,
    align=center,
    line width=0.8pt,
  },
  leaf/.style={
    my-box,
    minimum height=1.5em,
    text=black,
    align=center,
    font=\normalsize,
    inner xsep=2pt,
    inner ysep=4pt,
    line width=0.8pt,
  }
}

\begin{document}
%\setcopyright{none}

\title{Reconciling Methodological Paradigms: Employing Large Language Models as Novice Qualitative Research Assistants in Talent Management Research 
}


\author{Sreyoshi Bhaduri}
\affiliation{%
\institution{Amazon}
\city{Arlington}
\state{Virginia}
\country{USA}}

\author{Satya Kapoor}
\affiliation{%
\institution{Amazon}
\city{Vancouver}
\state{British Columbia}
\country{Canada}}

\author{Alex Gil}
\affiliation{%
\institution{Amazon}
\city{Arlington}
\state{Virginia}
\country{USA}}

\author{Anshul Mittal}
\affiliation{%
\institution{Amazon}
\city{Arlington}
\state{Virginia}
\country{USA}}

\author{Rutu Mulkar}
\affiliation{%
\institution{Amazon}
\city{Seattle}
\state{Washington}
\country{USA}}

\renewcommand{\shortauthors}{Accepted to Talent Management and Computing (TMC) Workshop at KDD '24 at Barcelona, Spain}


%%%%%%%%%%%%%%%%%%
%paper begins here
%%%%%%%%%%%%%%%%%%
\begin{abstract}

Qualitative data collection and analysis approaches, such as those employing interviews and focus groups, provide rich insights into customer attitudes, sentiment, and behavior. However, manually analyzing qualitative data requires extensive time and effort to identify relevant topics and thematic insights. This study proposes a novel approach to address this challenge by leveraging Retrieval Augmented Generation (RAG) based Large Language Models (LLMs) for analyzing interview transcripts. The novelty of this work lies in strategizing the research inquiry as one that is augmented by an LLM that serves as a novice research assistant. This research explores the mental model of LLMs to serve as novice qualitative research assistants for researchers in the talent management space. A RAG-based LLM approach is extended to enable topic modeling of semi-structured interview data, showcasing the versatility of these models beyond their traditional use in information retrieval and search. Our findings demonstrate that the LLM-augmented RAG approach can successfully extract topics of interest, with significant coverage compared to manually generated topics from the same dataset. This establishes the viability of employing LLMs as novice qualitative research assistants. Additionally, the study recommends that researchers leveraging such models lean heavily on quality criteria used in traditional qualitative research to ensure rigor and trustworthiness of their approach. Finally, the paper presents key recommendations for industry practitioners seeking to reconcile the use of LLMs with established qualitative research paradigms, providing a roadmap for the effective integration of these powerful, albeit novice, AI tools in the analysis of qualitative datasets within talent management research.

\end{abstract}



\maketitle
%\pagestyle{empty}

%%%%%%%%%%%%%%%%%%
%paper begins here
%%%%%%%%%%%%%%%%%%

\section{Introduction}
Talent management researchers frequently work backwards from their customers, the employees at the organization. Understanding employee sentiment and behavior often involves conducting deep-dive interviews, explanatory in nature – e.g., demystifying the why behind customer choices, attitudes or behaviors (e.g., \cite{leino2007case}). Talent management research, at its core, seeks to use science to equip every employee with resources to help them best navigate their careers \cite{zhao2023using}.

Consequently, qualitative research methodology plays a critical role in talent management. Many of the key considerations around employee engagement, motivation, and workforce culture involve subjective, context-dependent factors that are best explored through in-depth interviews, focus groups, and other qualitative data collection approaches. Talent management professionals often rely on rich qualitative datasets to gain deep insights into employee experiences, organizational dynamics, and the nuances of human capital. However, these qualitative paradigms can clash with the more positivist, quantitative worldview that underlies many of the analytic tools used to evaluate talent management data. Talent management researchers may find that standard statistical techniques and data visualization approaches struggle to fully capture the complexities inherent in qualitative datasets, leading to potential misinterpretations or oversimplifications of the human elements involved in managing an organization's workforce. Navigating this tension between qualitative and quantitative approaches is an ongoing challenge for talent management professionals. 

Large language models (LLMs) like BERT, GPT-3 and PaLM have demonstrated strong aptitude for summarization (e.g., \cite{yang2023recent}), classification (e.g., \cite{pelaez2024large}), and information extraction (e.g., \cite{dunn2022structured}) for text-based data. Consequently, LLMs are also increasingly being leveraged within talent management contexts for tasks such as interview analysis. However, language models are themselves designed primarily from a quantitative, data-driven paradigm. These models are trained on vast troves of text data using statistical machine learning techniques optimized for numerical patterns and correlations. While powerful at extracting insights from large-scale datasets, LLMs can often struggle to fully capture the nuanced, contextual nature of language \cite{bender2021dangers}, \cite{dwivedi2023opinion} that is critical for qualitative information sourced from interviews, focus groups, and other qualitative research methods common in talent management.

Talent management professionals must therefore continuously navigate a tension between the quantitative orientation of their analytical tools and the qualitative richness of the human dynamics they seek to understand. Bridging this gap requires innovative approaches that combine the opportunity for scale and speed offered by LLM-powered analysis augmented by borrowing evaluative nuances of traditional qualitative techniques. Talent leaders, thus, must carefully select and configure their AI-powered tools to ensure the voices and experiences of employees are authentically represented, rather than reduced to oversimplified metrics. Mastering this balance is an ongoing challenge, but one that is critical for talent management to yield truly holistic and impactful insights.

This paper presents results from leveraging LLMs as a novice qualitative researcher to augment qualitative research workstreams, specifically for data generated through semi-structured interviews.

The purpose of this paper is two-fold – 1) provide an overview of a successful implementation of a Retrieval Augmented Generation-based model for analyzing semi-structured interviews, and more importantly, 2) enumerate pragmatic take-aways and learnings drawing from traditional qualitative research to help fellow industry practitioners in reconciling the methodological paradigms. We posit the second purpose to be valuable to the larger discussion within talent management research communities on how and where to integrate AI capabilities across different talent management workstreams. 

%%%%%%%%%%%%%%%%%%
%paper intro ends
%%%%%%%%%%%%%%%%%%

\section {Quantitative and Qualitative Paradigms} 
Quantitative and qualitative research represent two fundamental paradigms or philosophical frameworks that guide research strategies, methods, analysis, and use of results \cite{yilmaz2013comparison}. While both methodological approaches seek to rigorously study research problems, they are based on distinct assumptions and procedures adapted to investigating particular types of questions and drawing different conclusions. Quantitative research is based on the assumptions of positivism, the philosophical tradition premised on the application of natural science methods to the study of social reality and beyond \cite{bryman2016social}. Quantitative researchers believe that objective facts and truths about human behavior and society can be measured and quantified numerically. Quantitative methods such as surveys, structured observations, and experiments aim to test hypotheses derived from theories by examining relationships between precisely measured variables statistically analyzed using large sample sizes \cite{creswell2017research}. These methods seek to minimize subjectivity and generalize findings to a population. In contrast, qualitative research aligns with interpretivist and constructivist philosophical traditions by embracing subjectivity and focused meaning-making by and with research participants \cite{denzin2023sage}. 

Qualitative researchers often use an inductive approach aimed at discovering and understanding processes, experiences, and worldviews by collecting non-numerical data through methods like in-depth interviews, ethnographic fieldwork, and document analysis. Findings derive from themes that emerge openly from the data rather than testing predetermined hypotheses. Samples tend to be small and purposely selected to illuminate a phenomenon in depth and detail. The aim is particularization rather than generalization, with a priority on ecological validity and multiple realities situated in time, place, culture, and context. 

While debates once positioned these paradigms in opposition, contemporary mixed methods research leverages the complementary strengths of quantitative and qualitative approaches \cite{halcomb2015mixed}. Mixed methods investigations integrate quantitative and qualitative data collection and analysis within a single program of inquiry by combining these approaches in creative ways to deepen understanding \cite{creamer2017introduction} \cite{creamer2018striving} \cite{greene2008mixed}. This reconciliation of methodological perspectives offers opportunities to generate more robust, contextualized insights to address complex research problems. The use of large language models (LLMs) as novice qualitative research assistants, as explored in this paper, can be considered an exercise in mixed methods research design. 

Prior to LLMs, in previous work, Natural Language Processing based modeling of qualitative data from social science contexts, have also been used as "novice insight" augmented by the more expert contextualization provided by human researchers (e.g., \cite{bhaduri2018nlp}, \cite{bhaduri2021semester}). Popular traditional topic modeling techniques (e.g. Latent Dirichlet Allocation), however, suffer from several limitations (e.g. specifying number of clusters) when compared to existing deep learning-based methods. They also often fail to capture the contextual nuances and ambiguities inherent in natural language, as they rely heavily on predefined rules and patterns \cite{devlin2018bert} \cite{radford2019language}. This can make it challenging to handle the complexities and variations present in real-world text data, and may require domain-specific knowledge or fine-tuning to achieve acceptable performance \cite{lee2019patentbert}. Recent advancements in LLMs, such as BERT and GPT, have largely overcome these limitations by leveraging deep neural networks to learn rich, contextual representations from large amounts of text data \cite{NIPS2017_3f5ee243} \cite{devlin2018bert}. These powerful models can capture subtle semantic and pragmatic features of language, and demonstrate strong generalization capabilities through transfer learning \cite{brown2020language} \cite{radford2019language}. 

Further, in traditional qualitative research, thematic analysis is the process of gathering themes across topics from qualitative data, such as interview data, through iteratively analyzing the dataset for topics of interest \cite{creamer2017introduction}. Inductive coding and deductive coding are two approaches to analyzing data from semi-structured interviews. Inductive coding involves starting with raw data and gradually developing codes and categories based on patterns and topics that emerge from the data as the researcher manually interacts with it \cite{patton2014qualitative} \cite{strauss1998basics}. This approach is bottom-up, where the data drives the development of codes and theories \cite{glaser1965constant}. Deductive coding, on the other hand, involves starting with preconceived codes or theories and applying them to the data \cite{pearse2019illustration}. This approach is top-down, where existing theories or frameworks guide the coding process \cite{maxwell2018collecting}. Researchers in industry typically work backwards from research question of interest. Most of the research questions in industry driving qualitative data collection are also explanatory (i.e., tend to explain the quantitative findings such as low customer satisfaction, low product adoption numbers), rather than exploratory (i.e., ethnography of a community of interest or a phenomenon) and as a result deductive approaches are often more popular than inductive coding.

Ultimately, by augmenting traditional deep-dive qualitative analysis with the time and resource efficient pattern recognition and text processing capabilities of LLMs, researchers can integrate quantitative and qualitative techniques to enhance the speed, depth, and rigor of their investigations. This mental model of a novice-LLM approach holds promise for bridging the divide between positivist and interpretive paradigms, ultimately working towards a more comprehensive understanding of the phenomenon under study.
%%%%%%%%%%%%%%%%%%
%paper background ends
%%%%%%%%%%%%%%%%%%

%%%%%%%%%%%%%%%%%%
%figure begins here
%%%%%%%%%%%%%%%%%%
\begin{figure*}
   \centering
    \resizebox{\textwidth}{!}{%
    \includegraphics{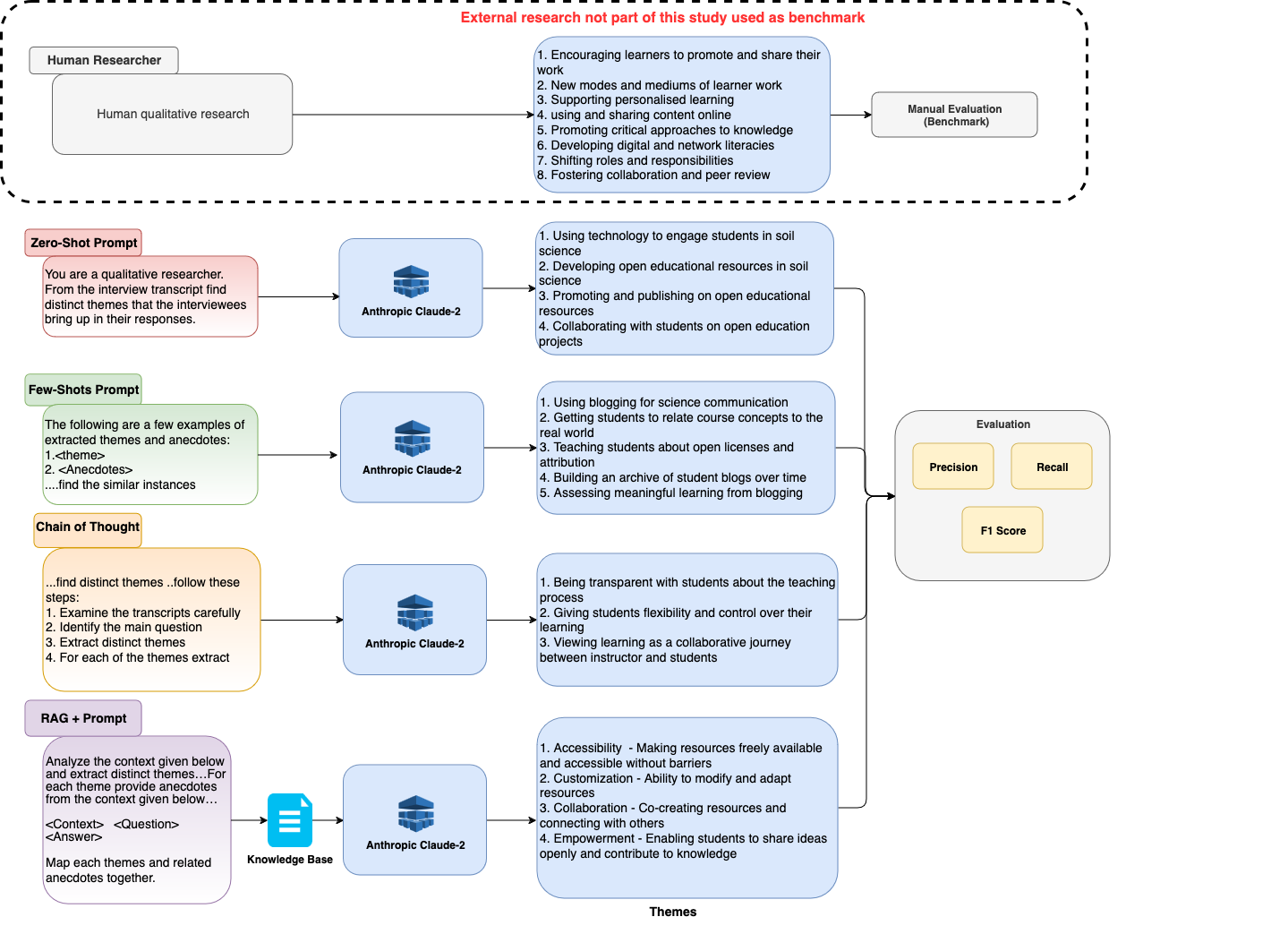}}
    \caption{Comparison across prompting approaches}
    \label{fig:SSA}
\end{figure*}
%%%%%%%%%%%%%%%%%%
%figure ends here
%%%%%%%%%%%%%%%%%%

\section{Dataset}
We used an open-source dataset \cite{paskevicius2018exploring} to demonstrate how an LLM prompted as a novice researcher can enhance traditional qualitative deductive thematic coding. This dataset was originally collected to explore educators’ experiences implementing open educational practices \cite{paskevicius2018exploring}. The dataset contains eight transcripts each from hour-long interviews conducted with educators to understand how they are using openly accessible sources of knowledge and open-source tools. The original research involved a deep-dive qualitative analysis through using a phenomenological approach to extract topics manually from the dataset. We chose this open-source dataset for two reasons – 1) structural match to proprietary dataset, and 2) rich description and manually identified topics by an expert to serve as a gold standard to measure the efficacy of our LLM based approach. Semi-structured interviews provide critical insights through participant perspectives, making them foundational in various industry settings. 

The semi-structured approach used to create this dataset is a close match to proprietary talent management data from our organization, where employees are interviewed on a particular phenomenon to get deeper understanding of their related sentiment, attitudes, and behaviors. Manually extracted topics serve as gold standard for benchmarking findings from our LLM-based approach. The paper \cite{paskevicius2018exploring} describing the dataset explains the manual process establishing how each transcript was read twice: first, for a comprehensive analysis, and subsequently, to initiate a thematic exploration. Additional reviewing continued as codes and topics emerged and intersected among the interviews. A manual qualitative coding approach was applied at each iteration to reveal themes, following constant comparison methodology \cite{glaser1965constant}. 

We posit that our approach, as demonstrated on this sample semi-structured interview dataset, can easily extend to multiple industry settings in talent management research where researchers conduct interviews and focus groups.

%%%%%%%%%%%%%%%%%%
%figure begins here
%%%%%%%%%%%%%%%%%%
\begin{figure*}
   \centering
    \resizebox{\textwidth}{!}{%
    \includegraphics{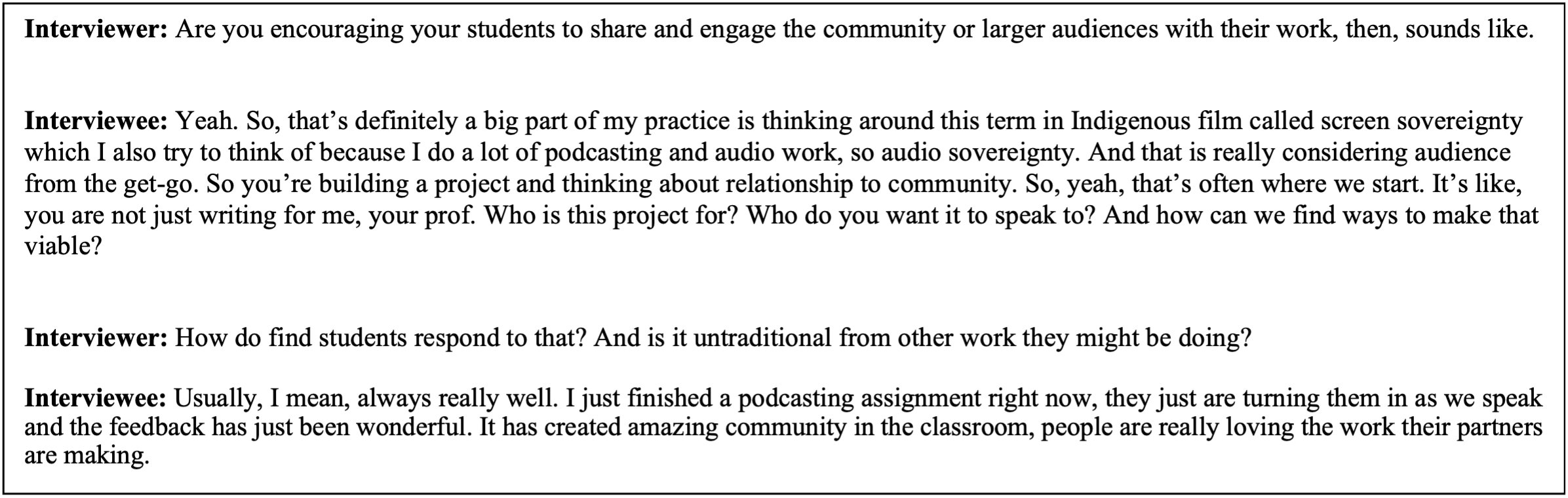}}
    \caption{Sample of the interview transcript}
\end{figure*}
%%%%%%%%%%%%%%%%%%
%figure ends here
%%%%%%%%%%%%%%%%%%

\section{Thematic Analysis Using LLMs}
In traditional, manual qualitative research, deductive thematic analysis process begins with the researcher first formulating the research questions. Then, upon collection of the data, such as interview transcripts, the researcher iterates manually through the transcripts to identify and extract themes or topics of interest. This labor-intensive process involves carefully reading through the data, taking notes, and organizing the topics iteratively into broader coherent themes that address the research questions. The researcher may go through multiple rounds of coding and analysis to refine the themes and ensure they comprehensively capture the key insights from the data. Our approach finds that LLMs can quickly uncover topics of interest from the dataset which can then be iterated upon to garner broader themes 
 of interest across topics. Thus, for our novice-LLM led approach, we leveraged the power of Large Language Models (LLMs) as a novice research assistant in the thematic analysis process. Specifically, we used the open-source framework called Langchain to create dynamic prompt templates, such as few-shot prompts and chain of thoughts, that guided the LLM in performing topic modeling and generating insights from the interview transcripts. We then opted to use Anthropic's Claude2 model to execute these prompts and extract the relevant themes.

To initiate the analysis, we first selected a main research question and corresponding sub-questions from our dataset \cite{paskevicius2018exploring}. We then fed these research questions, along with the interview transcripts, into the LLM-powered Langchain framework. The model was able to quickly identify and summarize the key topics, and iteratively, themes emerging from the data.  This approach provided a quick yet relatively comprehensive analysis that would have taken a human researcher significant time and effort to reproduce manually. 

\subsection{Thematic analysis enhanced through Retrieval Augmented Generation (RAG)}

In our LLM based approaches, we experiment with four methods - zero-shot prompting, few-shot prompting, chain-of-thought reasoning, and Retrieval Augmented Generation based Question Answering. In zero-shot prompting we provide a single prompt to the model. In few-shot prompting, we provide a set of topics and anecdotes to the model as examples. In the chain of thought (COT) approach, we provide a set of instructions for the model to follow. Finally, for Retrieval Augmented Generation (RAG) we provide context and questions to the model, from which it extracts information. 

Zero-shot prompts are simple instructions or tasks given to an LLM that have not been specifically trained on that task. It serves as a baseline because it demonstrates the model's fundamental ability to understand and respond to prompts based solely on its pre-training \cite{kong2023better}. In few-shot prompting, a small set of examples illustrating the desired outcome are manually selected and provided to the LLM. These examples allow the model to understand the tasks at hand and generate similar results \cite{brown2020language}. Chain-of-thought prompting provides a set of intermediate steps to guide the LLM to mimic human-like reasoning. This significantly improves the capability of the LLM to understand complex reasoning and generate better topics \cite{weizhou}.  Retrieval-augmented generation (RAG) combines the capabilities of an LLM with a retrieval system to source and integrate additional information into its responses \cite{lewis2020retrieval}. This effort provides contextually richer and ultimately more accurate outputs. We do this by providing all the interview transcripts to the LLM as a custom knowledge base. Two considerations helped the RAG approach outperform the other approaches:

\subsubsection{Focused Analysis:} In our approach, LLM searches the knowledge base to find and retrieve parts of documents that are most relevant to the question in the query. This narrows the focus to the most relevant information and ensures attention to critical topics and nuances.

\subsubsection{Context Dilution/Managing Information Overload:} Using all transcripts as input in a single instance creates information overload scenarios, ultimately leading to dilution of important topics or nuances. If the dataset is too large or complex, LLM might lose track of what’s most relevant to specific query, leading hallucinations.  Hallucinations or inaccuracies within this context refers to instances where the model generates information which is not grounded in input data. In our approach, the use of RAG mitigates some of the hallucination by anchoring LLM responses relevant information, and providing a form of contextual validation for the output. 

%%%%%%%%%%%%%%%%%%
%paper Methodology ends
%%%%%%%%%%%%%%%%%%

%%%%%%%%%%%%%%%%%%
%paper Findings 
%%%%%%%%%%%%%%%%%%

\begin{table*}[h!]
    \centering
    \normalsize
    \begin{tabular}{cccc}
        \hline
        \\ \textbf{Distillbert-base-uncased}
         & Precision & Recall & F1-Score \\
        \hline
        Chain of Thought & 67\% & 62\% & 64\% \\
        Few Shot & 72\% & 67\% & 70\% \\
        Zero Shot & 68\% & 66\% & 67\% \\
        \textbf{RAG} & \textbf{79\%} & \textbf{80\%} & \textbf{79\%}\\
        \hline
        
        \hline
        \\ \textbf{Bert-base-uncased}
         & Precision & Recall & F1-Score \\
        \hline
        Chain of Thought & 56\% & 48\% & 52\% \\
        Few Shot & 64\% & 56\% & 60\% \\
        Zero Shot & 59\% & 55\% & 57\% \\
        \textbf{RAG} & \textbf{70\%} & \textbf{70\%} & \textbf{70\%}\\

        \hline
        \\ \textbf{Roberta-large}
         & Precision & Recall & F1-Score \\
        \hline
        Chain of Thought & 89\% & 85\% & 87\% \\
        Few Shot & 90\% & 87\% & 88\% \\
        Zero Shot & 89\% & 86\% & 88\% \\
        \textbf{RAG} & \textbf{92\%} & \textbf{91\%} & \textbf{91\%}\\
        \\
        \hline
        \\
    \end{tabular}
    \caption{Comparison of Results across the LLM Enhanced Thematic Analysis Strategies Employed}
    \label{tab:table-label}
\end{table*}

\section{Findings}
In the paper describing the dataset leveraged for this work, the authors collected and conducted a manual analysis \cite{paskevicius2018exploring}. Their research led to identification of significant, recurring topics within the interviews. Our evaluation strategy uses these manually generated topics from the paper's work as gold standard to compare against topics generated by the LLMs-based approach. We use Precision (\equationautorefname{ 1}), Recall (\equationautorefname{ 2}), and F1-score (\equationautorefname{ 3}) to benchmark topics generated by our LLM-augmented qualitative research approach against the topics generated by the human researcher.

\begin{equation}
\label{eq:1}
\text{R}_{BERT} = \frac{1}{|x|} \sum_{x_i \in x} \max_{x_j \in \hat{x}} x_i^T \hat{x}_j
\end{equation}

\begin{equation}
\label{eq:2}
\text{P}_{BERT} = \frac{1}{|\hat{x}|} \sum_{\hat{x}_i \in \hat{x}} \max_{x_j \in x} \hat{x}_i^T x_j
\end{equation}

\begin{equation}
\label{eq:3}
\text{F}_{BERT} = 2 \cdot \frac{P_\text{BERT} \cdot R_\text{BERT}}{P_\text{BERT} + R_\text{BERT}}
\end{equation}

\begin{equation}
\label{eq:4}
\text{Cosine Similarity} = \frac{x_i^T \hat{x}_j}{\|x_i\| \|\hat{x}_j\|}
\end{equation}

These metrics are the current evaluation standard for classification models, but they can be adapted for text generation tasks \cite{zhang2019bertscore}.  Precision and Recall measure the proportion of correctly identified positive cases. In the context of our experiment, every word from predicted text gets matched to a word in the referenced text to compute recall. This process is inverted to then compute precision. The precision and recall values are then combined to compute an F1 score. These metrics use cosine similarity (\equationautorefname{ 4}) in which each predicted word is paired with its closest corresponding word from the reference text with the aim of maximizing the similarity score.

In Table 1, the performance of various LLM prompting techniques including Chain of Thought, Few Shot, Zero Shot and RAG, are compared across different embedding models (Distillbert-base-uncased, Bert-base-uncased, and Roberta-large). This comparison aims to evaluate the robustness and effectiveness of these prompting techniques. Our results indicate that while each prompting technique shows varying level of precision, recall and F1-score, RAG consistently outperform the others on all three metrics, achieving highest performance across all models.

%%%%%%%%%%%%%%%%%%
%paper Findings ends
%%%%%%%%%%%%%%%%%%
\begin{table*}[h!]
    \centering
    \normalsize
    \begin{tabular}{cc}
        \hline
        \\Example: Keywords from LDA Topic One \\
        \\
        Students\\ Course \\
        Develop\\ People \\
        Institution\\ Project \\
        Science\\ Discipline \\
        Material\\ Start \\
        \\
        \hline
        \\
        \\Example: Output from LLM approach\\
        \\
        Collaboration: Co-creating resources and connecting with others \\ 
        \\
        \textit{Corresponding Anecdote: You can also in your teaching have students connect with people outside the} 
        \\\textit{course in various ways. Like, maybe some people outside the course are commenting on blogs}\\ \textit{and student are getting in a conversation around that.}\\
        \\
    \hline
        \\
    \end{tabular}
    \caption{Example Output from LLM approach compared to Keywords from LDA Topic One }
    \label{tab:table-label}
\end{table*}

\section{Learnings}
Treating large language models (LLMs) as novice research assistants during thematic analysis offered valuable insights for our research. By framing the LLM as a novice collaborator with little knowledge or insight of the context, prompts can be crafted to better guide the model and leverage its capabilities. Used prudently, similar novice LLM-augmented approaches can significantly increase time and resource efficiency compared to traditional qualitative coding methods in talent management research. The following sections explore some of our key learnings that may benefit other researchers considering designing LLMs as novice researchers to optimize thematic analysis.

\subsection{Approaching LLMs as Novice Research Assistants can help prepare better prompts}

A novice is a person who, “has no experience with the situations in which they are expected to perform tasks” \cite{benner1982novice}. The novice is thus at a basic proficiency level for skill acquisition, with limited information and prior experience related to a task at hand \cite{montfort2013novice}. For large qualitative datasets analyzed using LLMs we propose that a novice-led approach to analysis is a good fit. In our approach the human behaves as an expert prompting the novice LLM to provide insights related to topics of interest. We found this framework as a helpful mental model to ground the primary researcher  prompting the LLM as they iteratively uncover insights from the dataset.

\subsection{Used prudently, LLMs can help increase time effectiveness and resource efficiency}

LLMs have advanced the field of natural language processing with their ability to understand and generate responses that closely mimic human language \cite{shanahan2024talking}. The strengths of LLMs extend beyond metrics, these models are adept at processing vast amounts of text rapidly, demonstrating a level of topic modeling that can mimic human analysis. Manual topic modeling is human labor intensive and time inefficient \cite{clarke2017thematic}. LLMs also enhance efficiency by streamlining the processing of large datasets, allowing for the extraction of topics from qualitative data more quickly. Improvisations of these model using techniques like few-shot and zero-shot learning capabilities further reduce the need for expensive data labeling and annotations. In a nutshell, LLMs boost speed, reduce human effort, scale to massive datasets, and lower labeling costs. However, human expertise is still essential for judgment, validation and end-to-end framework design.

\subsection{LLM augmented approaches offer significant increase in ease and enhanced context compared to traditional NLP approaches.} Using a RAG approach towards an LLM-augmented qualitative research analyzing semi-structure interviews shows great promise compared to natural language processing methods like Latent Dirichlet allocation (LDA). Currently, there are no widely accepted methods for comparing the two approaches as there is no bridge to compare keywords to themes, except from a human-evaluator ease of interpretability standpoint. We performed topic modeling analysis on the same dataset with the broader aim of finding themes. Manually comparing both approaches, each researcher of this workstream independently found that any of the approaches using an LLM yielded much greater context and consequently, better interpretability than the traditional LDA approach. This is likely because, with LDA, the model outputs a list of words and probability for each topic. With these words, the researcher would then have to manually define the topic. While this approach increases researcher flexibility, it remains time and resource consuming. In contrast, with the LLM approach, the output is richer in context of what particular topics mean. For example, our LDA model yielded 5 topics (see: Appendix A \figureautorefname{ 3}). The first 10 words for topic 1 can also be seen in Table 2. Putting these words together into a comprehensive theme can be challenging without more context. However, an LLM is able to generate context grounded in the participant’s voice for researchers to work with. An example of an extracted theme and its corresponding anecdote using an LLM can also be seen in Table 2, above.

%%%%%%%%%%%%%%%%%%
%paper Summary
%%%%%%%%%%%%%%%%%%
\section{Recommendations}
Traditional qualitative research is evaluated based on several criteria that ensure quality and rigor of the research, both in terms of methods as well as findings. Prior research has established four criteria for increased rigor and trustworthiness of qualitative research studies around credibility, dependability, confirmability, and transferability \cite{lincoln1988criteria}. We recommend three ways in which quality criteria from traditional qualitative research can be used by practitioners employing LLM augmented analysis of qualitative data.

\subsection{Establishing credibility of findings by incorporating mechanism for member checks.} Member checks, i.e., the strategy of soliciting insights from research participants on research findings, are often relied on as the gold standard for increasing trustworthiness of qualitative research approaches (e.g., \cite{patton2014qualitative} \cite{kornbluh2015combatting}). Qualitative researchers employing LLMs can work on deepening their understanding of the research context using appropriate data-collection methods and tools that work best for particular contexts, as well as conduct adequate member checking to ensure the accuracy of findings.   

\subsection{Practicing increased researcher reflexivity.} Qualitative researchers are recommended that they acknowledge and address their own biases, thus recognizing the influence of their own experiences and opinions on the research process \cite{finlay2002negotiating}. Similar exercises on reflectivity can also be helpful for researchers augmenting qualitative data analysis through employing LLMs. Researcher reflexivity in such instances can extend to querying the LLM to ask for rationale on why certain topics were extracted, grounding topics in anecdotes from the transcripts, and recognizing the influence the human researcher’s prior knowledge and biases will have on the prompts used. Future work in extending LLMs for qualitative research should continue to draw on evaluation criteria grounded in traditional qualitative research paradigm.

\subsection{Increasing transparency of decisions made throughout the research study.}Qualitative researchers are recommended to thoroughly document all decisions that guide their analysis process by providing thick descriptions, allowing for increased transparency. This practice enhances reliability and reproducibility of the research \cite{lincoln1988criteria}. Qualitative researchers employing LLMs should also similarly strategize maximizing transparency through mechanisms such as documenting changes in workflow, sharing prompts, and detailing model preferences.

%%%%%%%%%%%%%%%%%%
%paper Summary ends
%%%%%%%%%%%%%%%%%%

\section{Closing Thoughts}
The approach outlined in this paper offers a promising avenue for industry-based talent management practitioners seeking to increase the time and resource efficiency of qualitative interview data analysis. By leveraging large language models (LLMs) as novice qualitative research assistants, organizations can potentially accelerate the coding, categorization, and thematic synthesis of rich interview data - a critical bottleneck in many talent management research initiatives.

However, as the field of LLM-assisted qualitative research matures, it will be essential to not only benchmark model performance against traditional quantitative evaluation metrics, but also consider quality criteria more prominent within the qualitative research paradigm. Factors such as credibility, transferability, dependability, and confirmability will need to be carefully evaluated as LLMs are integrated into qualitative workflows. Furthermore, the ethical use of AI assistants in sensitive domains like talent management will require close, multi-disciplinary attention to issues at the intersection of data privacy, algorithmic bias, and model transparency, for which researchers will have to be trained \cite{mackenzie2024beyond}. 

Future research should seek to establish guidelines and best practices for LLM-augmented qualitative analysis that uphold the rigor and trustworthiness expected within the qualitative research community. Only by doing so can talent management scholars and practitioners unlock the full potential of these powerful language models, while respecting the epistemological foundations of qualitative inquiry. As the field evolves, we believe that a judicious, ethically-grounded approach to LLM integration can yield substantial gains in research efficiency and organizational impact.

%%%%%%%%%%%%%%%%%%
%paper Summary ends
%%%%%%%%%%%%%%%%%%

\bibliographystyle{ACM-Reference-Format}
\bibliography{sample-base}


\begin{thebibliography}{40}

%%% the Web address.

\ifx \showCODEN    \undefined \def \showCODEN     #1{\unskip}     \fi
\ifx \showDOI      \undefined \def \showDOI       #1{#1}\fi
\ifx \showISBNx    \undefined \def \showISBNx     #1{\unskip}     \fi
\ifx \showISBNxiii \undefined \def \showISBNxiii  #1{\unskip}     \fi
\ifx \showISSN     \undefined \def \showISSN      #1{\unskip}     \fi
\ifx \showLCCN     \undefined \def \showLCCN      #1{\unskip}     \fi
\ifx \shownote     \undefined \def \shownote      #1{#1}          \fi
\ifx \showarticletitle \undefined \def \showarticletitle #1{#1}   \fi
\ifx \showURL      \undefined \def \showURL       {\relax}        \fi
% invisible to TeX
\providecommand\bibfield[2]{#2}
\providecommand\bibinfo[2]{#2}
\providecommand\natexlab[1]{#1}
\providecommand\showeprint[2][]{arXiv:#2}

\bibitem[Bender et~al\mbox{.}(2021)]%
        {bender2021dangers}
\bibfield{author}{\bibinfo{person}{Emily~M Bender}, \bibinfo{person}{Timnit Gebru}, \bibinfo{person}{Angelina McMillan-Major}, {and} \bibinfo{person}{Shmargaret Shmitchell}.} \bibinfo{year}{2021}\natexlab{}.
\newblock \showarticletitle{On the dangers of stochastic parrots: Can language models be too big?}. In \bibinfo{booktitle}{\emph{Proceedings of the 2021 ACM conference on fairness, accountability, and transparency}}. \bibinfo{pages}{610--623}.
\newblock


\bibitem[Benner(1982)]%
        {benner1982novice}
\bibfield{author}{\bibinfo{person}{Patricia Benner}.} \bibinfo{year}{1982}\natexlab{}.
\newblock \showarticletitle{From novice to expert}.
\newblock \bibinfo{journal}{\emph{AJN The American Journal of Nursing}} \bibinfo{volume}{82}, \bibinfo{number}{3} (\bibinfo{year}{1982}), \bibinfo{pages}{402--407}.
\newblock


\bibitem[Bhaduri(2018)]%
        {bhaduri2018nlp}
\bibfield{author}{\bibinfo{person}{Sreyoshi Bhaduri}.} \bibinfo{year}{2018}\natexlab{}.
\newblock \showarticletitle{NLP in Engineering Education-Demonstrating the use of Natural Language Processing Techniques for Use in Engineering Education Classrooms and Research}.
\newblock  (\bibinfo{year}{2018}).
\newblock


\bibitem[Bhaduri et~al\mbox{.}(2021)]%
        {bhaduri2021semester}
\bibfield{author}{\bibinfo{person}{Sreyoshi Bhaduri}, \bibinfo{person}{Michelle Soledad}, \bibinfo{person}{Tamoghna Roy}, \bibinfo{person}{Homero Murzi}, {and} \bibinfo{person}{Tamara Knott}.} \bibinfo{year}{2021}\natexlab{}.
\newblock \showarticletitle{A Semester Like No Other: Use of Natural Language Processing for Novice-Led Analysis on End-of-Semester Responses on Students’ Experience of Changing Learning Environments Due to COVID-19}. In \bibinfo{booktitle}{\emph{2021 ASEE Virtual Annual Conference Content Access}}.
\newblock


\bibitem[Brown(2020)]%
        {brown2020language}
\bibfield{author}{\bibinfo{person}{Tom~B Brown}.} \bibinfo{year}{2020}\natexlab{}.
\newblock \showarticletitle{Language models are few-shot learners}.
\newblock \bibinfo{journal}{\emph{arXiv preprint ArXiv:2005.14165}} (\bibinfo{year}{2020}).
\newblock


\bibitem[Bryman(2016)]%
        {bryman2016social}
\bibfield{author}{\bibinfo{person}{Alan Bryman}.} \bibinfo{year}{2016}\natexlab{}.
\newblock \bibinfo{booktitle}{\emph{Social research methods}}.
\newblock \bibinfo{publisher}{Oxford university press}.
\newblock


\bibitem[Clarke and Braun(2017)]%
        {clarke2017thematic}
\bibfield{author}{\bibinfo{person}{Victoria Clarke} {and} \bibinfo{person}{Virginia Braun}.} \bibinfo{year}{2017}\natexlab{}.
\newblock \showarticletitle{Thematic analysis}.
\newblock \bibinfo{journal}{\emph{The journal of positive psychology}} \bibinfo{volume}{12}, \bibinfo{number}{3} (\bibinfo{year}{2017}), \bibinfo{pages}{297--298}.
\newblock


\bibitem[Creamer(2017)]%
        {creamer2017introduction}
\bibfield{author}{\bibinfo{person}{Elizabeth~G Creamer}.} \bibinfo{year}{2017}\natexlab{}.
\newblock \bibinfo{booktitle}{\emph{An introduction to fully integrated mixed methods research}}.
\newblock \bibinfo{publisher}{sage publications}.
\newblock


\bibitem[Creamer(2018)]%
        {creamer2018striving}
\bibfield{author}{\bibinfo{person}{Elizabeth~G Creamer}.} \bibinfo{year}{2018}\natexlab{}.
\newblock \bibinfo{title}{Striving for methodological integrity in mixed methods research: The difference between mixed methods and mixed-up methods}.
\newblock , \bibinfo{numpages}{526--530}~pages.
\newblock


\bibitem[Creswell and Creswell(2017)]%
        {creswell2017research}
\bibfield{author}{\bibinfo{person}{John~W Creswell} {and} \bibinfo{person}{J~David Creswell}.} \bibinfo{year}{2017}\natexlab{}.
\newblock \bibinfo{booktitle}{\emph{Research design: Qualitative, quantitative, and mixed methods approaches}}.
\newblock \bibinfo{publisher}{Sage publications}.
\newblock


\bibitem[Denzin et~al\mbox{.}(2023)]%
        {denzin2023sage}
\bibfield{author}{\bibinfo{person}{Norman~K Denzin}, \bibinfo{person}{Yvonna~S Lincoln}, \bibinfo{person}{Michael~D Giardina}, {and} \bibinfo{person}{Gaile~S Cannella}.} \bibinfo{year}{2023}\natexlab{}.
\newblock \bibinfo{booktitle}{\emph{The Sage handbook of qualitative research}}.
\newblock \bibinfo{publisher}{Sage publications}.
\newblock


\bibitem[Devlin(2018)]%
        {devlin2018bert}
\bibfield{author}{\bibinfo{person}{Jacob Devlin}.} \bibinfo{year}{2018}\natexlab{}.
\newblock \showarticletitle{Bert: Pre-training of deep bidirectional transformers for language understanding}.
\newblock \bibinfo{journal}{\emph{arXiv preprint arXiv:1810.04805}} (\bibinfo{year}{2018}).
\newblock


\bibitem[Dunn et~al\mbox{.}(2022)]%
        {dunn2022structured}
\bibfield{author}{\bibinfo{person}{Alexander Dunn}, \bibinfo{person}{John Dagdelen}, \bibinfo{person}{Nicholas Walker}, \bibinfo{person}{Sanghoon Lee}, \bibinfo{person}{Andrew~S Rosen}, \bibinfo{person}{Gerbrand Ceder}, \bibinfo{person}{Kristin Persson}, {and} \bibinfo{person}{Anubhav Jain}.} \bibinfo{year}{2022}\natexlab{}.
\newblock \showarticletitle{Structured information extraction from complex scientific text with fine-tuned large language models}.
\newblock \bibinfo{journal}{\emph{arXiv preprint arXiv:2212.05238}} (\bibinfo{year}{2022}).
\newblock


\bibitem[Dwivedi et~al\mbox{.}(2023)]%
        {dwivedi2023opinion}
\bibfield{author}{\bibinfo{person}{Yogesh~K Dwivedi}, \bibinfo{person}{Nir Kshetri}, \bibinfo{person}{Laurie Hughes}, \bibinfo{person}{Emma~Louise Slade}, \bibinfo{person}{Anand Jeyaraj}, \bibinfo{person}{Arpan~Kumar Kar}, \bibinfo{person}{Abdullah~M Baabdullah}, \bibinfo{person}{Alex Koohang}, \bibinfo{person}{Vishnupriya Raghavan}, \bibinfo{person}{Manju Ahuja}, {et~al\mbox{.}}} \bibinfo{year}{2023}\natexlab{}.
\newblock \showarticletitle{Opinion Paper:“So what if ChatGPT wrote it?” Multidisciplinary perspectives on opportunities, challenges and implications of generative conversational AI for research, practice and policy}.
\newblock \bibinfo{journal}{\emph{International Journal of Information Management}}  \bibinfo{volume}{71} (\bibinfo{year}{2023}), \bibinfo{pages}{102642}.
\newblock


\bibitem[Finlay(2002)]%
        {finlay2002negotiating}
\bibfield{author}{\bibinfo{person}{Linda Finlay}.} \bibinfo{year}{2002}\natexlab{}.
\newblock \showarticletitle{Negotiating the swamp: the opportunity and challenge of reflexivity in research practice}.
\newblock \bibinfo{journal}{\emph{Qualitative research}} \bibinfo{volume}{2}, \bibinfo{number}{2} (\bibinfo{year}{2002}), \bibinfo{pages}{209--230}.
\newblock


\bibitem[Glaser(1965)]%
        {glaser1965constant}
\bibfield{author}{\bibinfo{person}{Barney~G Glaser}.} \bibinfo{year}{1965}\natexlab{}.
\newblock \showarticletitle{The constant comparative method of qualitative analysis}.
\newblock \bibinfo{journal}{\emph{Social problems}} \bibinfo{volume}{12}, \bibinfo{number}{4} (\bibinfo{year}{1965}), \bibinfo{pages}{436--445}.
\newblock


\bibitem[Greene(2008)]%
        {greene2008mixed}
\bibfield{author}{\bibinfo{person}{Jennifer~C Greene}.} \bibinfo{year}{2008}\natexlab{}.
\newblock \showarticletitle{Is mixed methods social inquiry a distinctive methodology?}
\newblock \bibinfo{journal}{\emph{Journal of mixed methods research}} \bibinfo{volume}{2}, \bibinfo{number}{1} (\bibinfo{year}{2008}), \bibinfo{pages}{7--22}.
\newblock


\bibitem[Halcomb and Hickman(2015)]%
        {halcomb2015mixed}
\bibfield{author}{\bibinfo{person}{Elizabeth~J Halcomb} {and} \bibinfo{person}{Louise Hickman}.} \bibinfo{year}{2015}\natexlab{}.
\newblock \showarticletitle{Mixed methods research}.
\newblock  (\bibinfo{year}{2015}).
\newblock


\bibitem[Kong et~al\mbox{.}(2023)]%
        {kong2023better}
\bibfield{author}{\bibinfo{person}{Aobo Kong}, \bibinfo{person}{Shiwan Zhao}, \bibinfo{person}{Hao Chen}, \bibinfo{person}{Qicheng Li}, \bibinfo{person}{Yong Qin}, \bibinfo{person}{Ruiqi Sun}, {and} \bibinfo{person}{Xin Zhou}.} \bibinfo{year}{2023}\natexlab{}.
\newblock \showarticletitle{Better zero-shot reasoning with role-play prompting}.
\newblock \bibinfo{journal}{\emph{arXiv preprint arXiv:2308.07702}} (\bibinfo{year}{2023}).
\newblock


\bibitem[Kornbluh(2015)]%
        {kornbluh2015combatting}
\bibfield{author}{\bibinfo{person}{Mariah Kornbluh}.} \bibinfo{year}{2015}\natexlab{}.
\newblock \showarticletitle{Combatting challenges to establishing trustworthiness in qualitative research}.
\newblock \bibinfo{journal}{\emph{Qualitative research in psychology}} \bibinfo{volume}{12}, \bibinfo{number}{4} (\bibinfo{year}{2015}), \bibinfo{pages}{397--414}.
\newblock


\bibitem[Lee and Hsiang(2019)]%
        {lee2019patentbert}
\bibfield{author}{\bibinfo{person}{Jieh-Sheng Lee} {and} \bibinfo{person}{Jieh Hsiang}.} \bibinfo{year}{2019}\natexlab{}.
\newblock \showarticletitle{Patentbert: Patent classification with fine-tuning a pre-trained bert model}.
\newblock \bibinfo{journal}{\emph{arXiv preprint arXiv:1906.02124}} (\bibinfo{year}{2019}).
\newblock


\bibitem[Leino and R{\"a}ih{\"a}(2007)]%
        {leino2007case}
\bibfield{author}{\bibinfo{person}{Juha Leino} {and} \bibinfo{person}{Kari-Jouko R{\"a}ih{\"a}}.} \bibinfo{year}{2007}\natexlab{}.
\newblock \showarticletitle{Case amazon: ratings and reviews as part of recommendations}. In \bibinfo{booktitle}{\emph{Proceedings of the 2007 ACM conference on Recommender systems}}. \bibinfo{pages}{137--140}.
\newblock


\bibitem[Lewis et~al\mbox{.}(2020)]%
        {lewis2020retrieval}
\bibfield{author}{\bibinfo{person}{Patrick Lewis}, \bibinfo{person}{Ethan Perez}, \bibinfo{person}{Aleksandra Piktus}, \bibinfo{person}{Fabio Petroni}, \bibinfo{person}{Vladimir Karpukhin}, \bibinfo{person}{Naman Goyal}, \bibinfo{person}{Heinrich K{\"u}ttler}, \bibinfo{person}{Mike Lewis}, \bibinfo{person}{Wen-tau Yih}, \bibinfo{person}{Tim Rockt{\"a}schel}, {et~al\mbox{.}}} \bibinfo{year}{2020}\natexlab{}.
\newblock \showarticletitle{Retrieval-augmented generation for knowledge-intensive nlp tasks}.
\newblock \bibinfo{journal}{\emph{Advances in Neural Information Processing Systems}}  \bibinfo{volume}{33} (\bibinfo{year}{2020}), \bibinfo{pages}{9459--9474}.
\newblock


\bibitem[Lincoln and Guba(1988)]%
        {lincoln1988criteria}
\bibfield{author}{\bibinfo{person}{Yvonna~S Lincoln} {and} \bibinfo{person}{Egon~G Guba}.} \bibinfo{year}{1988}\natexlab{}.
\newblock \showarticletitle{Criteria for Assessing Naturalistic Inquiries as Reports.}
\newblock  (\bibinfo{year}{1988}).
\newblock


\bibitem[Mackenzie et~al\mbox{.}(2024)]%
        {mackenzie2024beyond}
\bibfield{author}{\bibinfo{person}{Tammy Mackenzie}, \bibinfo{person}{Leslie Salgado}, \bibinfo{person}{Sreyoshi Bhaduri}, \bibinfo{person}{Victoria Kuketz}, \bibinfo{person}{Solenne Savoia}, {and} \bibinfo{person}{Lilianny Virguez}.} \bibinfo{year}{2024}\natexlab{}.
\newblock \showarticletitle{Beyond the Algorithm: Empowering AI Practitioners through Liberal Education}. In \bibinfo{booktitle}{\emph{2024 ASEE Annual Conference \& Exposition}}.
\newblock


\bibitem[Maxwell(2018)]%
        {maxwell2018collecting}
\bibfield{author}{\bibinfo{person}{Joseph~A Maxwell}.} \bibinfo{year}{2018}\natexlab{}.
\newblock \showarticletitle{Collecting qualitative data: A realist approach}.
\newblock \bibinfo{journal}{\emph{The SAGE handbook of qualitative data collection}} (\bibinfo{year}{2018}), \bibinfo{pages}{19--32}.
\newblock


\bibitem[Montfort et~al\mbox{.}(2013)]%
        {montfort2013novice}
\bibfield{author}{\bibinfo{person}{Devlin~B Montfort}, \bibinfo{person}{Geoffrey~L Herman}, \bibinfo{person}{Shane~A Brown}, \bibinfo{person}{Holly~M Matusovich}, {and} \bibinfo{person}{Ruth~A Streveler}.} \bibinfo{year}{2013}\natexlab{}.
\newblock \showarticletitle{Novice-led paired thematic analysis: A method for conceptual change in engineering}. In \bibinfo{booktitle}{\emph{2013 ASEE Annual Conference \& Exposition}}. \bibinfo{pages}{23--933}.
\newblock


\bibitem[Paskevicius(2018)]%
        {paskevicius2018exploring}
\bibfield{author}{\bibinfo{person}{Michael Paskevicius}.} \bibinfo{year}{2018}\natexlab{}.
\newblock \emph{\bibinfo{title}{Exploring educators experiences implementing open educational practices}}.
\newblock \bibinfo{thesistype}{Ph.\,D. Dissertation}.
\newblock


\bibitem[Patton(2014)]%
        {patton2014qualitative}
\bibfield{author}{\bibinfo{person}{Michael~Quinn Patton}.} \bibinfo{year}{2014}\natexlab{}.
\newblock \bibinfo{booktitle}{\emph{Qualitative research \& evaluation methods: Integrating theory and practice}}.
\newblock \bibinfo{publisher}{Sage publications}.
\newblock


\bibitem[Pearse(2019)]%
        {pearse2019illustration}
\bibfield{author}{\bibinfo{person}{Noel Pearse}.} \bibinfo{year}{2019}\natexlab{}.
\newblock \showarticletitle{An illustration of deductive analysis in qualitative research}. In \bibinfo{booktitle}{\emph{18th European conference on research methodology for business and management studies}}. \bibinfo{pages}{264}.
\newblock


\bibitem[Pelaez et~al\mbox{.}(2024)]%
        {pelaez2024large}
\bibfield{author}{\bibinfo{person}{Sergio Pelaez}, \bibinfo{person}{Gaurav Verma}, \bibinfo{person}{Barbara Ribeiro}, {and} \bibinfo{person}{Philip Shapira}.} \bibinfo{year}{2024}\natexlab{}.
\newblock \showarticletitle{Large-scale text analysis using generative language models: A case study in discovering public value expressions in AI patents}.
\newblock \bibinfo{journal}{\emph{Quantitative Science Studies}} \bibinfo{volume}{5}, \bibinfo{number}{1} (\bibinfo{year}{2024}), \bibinfo{pages}{153--169}.
\newblock


\bibitem[Radford et~al\mbox{.}(2019)]%
        {radford2019language}
\bibfield{author}{\bibinfo{person}{Alec Radford}, \bibinfo{person}{Jeffrey Wu}, \bibinfo{person}{Rewon Child}, \bibinfo{person}{David Luan}, \bibinfo{person}{Dario Amodei}, \bibinfo{person}{Ilya Sutskever}, {et~al\mbox{.}}} \bibinfo{year}{2019}\natexlab{}.
\newblock \showarticletitle{Language models are unsupervised multitask learners}.
\newblock \bibinfo{journal}{\emph{OpenAI blog}} \bibinfo{volume}{1}, \bibinfo{number}{8} (\bibinfo{year}{2019}), \bibinfo{pages}{9}.
\newblock


\bibitem[Shanahan(2024)]%
        {shanahan2024talking}
\bibfield{author}{\bibinfo{person}{Murray Shanahan}.} \bibinfo{year}{2024}\natexlab{}.
\newblock \showarticletitle{Talking about large language models}.
\newblock \bibinfo{journal}{\emph{Commun. ACM}} \bibinfo{volume}{67}, \bibinfo{number}{2} (\bibinfo{year}{2024}), \bibinfo{pages}{68--79}.
\newblock


\bibitem[Strauss and Corbin(1998)]%
        {strauss1998basics}
\bibfield{author}{\bibinfo{person}{Anselm Strauss} {and} \bibinfo{person}{Juliet Corbin}.} \bibinfo{year}{1998}\natexlab{}.
\newblock \showarticletitle{Basics of qualitative research techniques}.
\newblock  (\bibinfo{year}{1998}).
\newblock


\bibitem[Vaswani et~al\mbox{.}(2017)]%
        {NIPS2017_3f5ee243}
\bibfield{author}{\bibinfo{person}{Ashish Vaswani}, \bibinfo{person}{Noam Shazeer}, \bibinfo{person}{Niki Parmar}, \bibinfo{person}{Jakob Uszkoreit}, \bibinfo{person}{Llion Jones}, \bibinfo{person}{Aidan~N Gomez}, \bibinfo{person}{\L~ukasz Kaiser}, {and} \bibinfo{person}{Illia Polosukhin}.} \bibinfo{year}{2017}\natexlab{}.
\newblock \showarticletitle{Attention is All you Need}. In \bibinfo{booktitle}{\emph{Advances in Neural Information Processing Systems}}, \bibfield{editor}{\bibinfo{person}{I.~Guyon}, \bibinfo{person}{U.~Von Luxburg}, \bibinfo{person}{S.~Bengio}, \bibinfo{person}{H.~Wallach}, \bibinfo{person}{R.~Fergus}, \bibinfo{person}{S.~Vishwanathan}, {and} \bibinfo{person}{R.~Garnett}} (Eds.), Vol.~\bibinfo{volume}{30}. \bibinfo{publisher}{Curran Associates, Inc.}
\newblock
\urldef\tempurl%
\url{https://proceedings.neurips.cc/paper_files/paper/2017/file/3f5ee243547dee91fbd053c1c4a845aa-Paper.pdf}
\showURL{%
\tempurl}


\bibitem[Wei et~al\mbox{.}({[n.\,d.]})]%
        {weizhou}
\bibfield{author}{\bibinfo{person}{J Wei}, \bibinfo{person}{X Wang}, \bibinfo{person}{D Schuurmans}, \bibinfo{person}{M Bosma}, \bibinfo{person}{F Xia}, {and} \bibinfo{person}{E Chi}.} \bibinfo{year}{[n.\,d.]}\natexlab{}.
\newblock \showarticletitle{\& Zhou, D.(2022)}.
\newblock \bibinfo{journal}{\emph{Chain-of-thought prompting elicits reasoning in large language models}} (\bibinfo{year}{[n.\,d.]}), \bibinfo{pages}{24824--24837}.
\newblock


\bibitem[Yang et~al\mbox{.}(2023)]%
        {yang2023recent}
\bibfield{author}{\bibinfo{person}{Binxia Yang}, \bibinfo{person}{Xudong Luo}, \bibinfo{person}{Kaili Sun}, {and} \bibinfo{person}{Michael~Y Luo}.} \bibinfo{year}{2023}\natexlab{}.
\newblock \showarticletitle{Recent progress on text summarisation based on bert and gpt}. In \bibinfo{booktitle}{\emph{International Conference on Knowledge Science, Engineering and Management}}. Springer, \bibinfo{pages}{225--241}.
\newblock


\bibitem[Yilmaz(2013)]%
        {yilmaz2013comparison}
\bibfield{author}{\bibinfo{person}{Kaya Yilmaz}.} \bibinfo{year}{2013}\natexlab{}.
\newblock \showarticletitle{Comparison of quantitative and qualitative research traditions: Epistemological, theoretical, and methodological differences}.
\newblock \bibinfo{journal}{\emph{European journal of education}} \bibinfo{volume}{48}, \bibinfo{number}{2} (\bibinfo{year}{2013}), \bibinfo{pages}{311--325}.
\newblock


\bibitem[Zhang et~al\mbox{.}(2019)]%
        {zhang2019bertscore}
\bibfield{author}{\bibinfo{person}{Tianyi Zhang}, \bibinfo{person}{Varsha Kishore}, \bibinfo{person}{Felix Wu}, \bibinfo{person}{Kilian~Q Weinberger}, {and} \bibinfo{person}{Yoav Artzi}.} \bibinfo{year}{2019}\natexlab{}.
\newblock \showarticletitle{Bertscore: Evaluating text generation with bert}.
\newblock \bibinfo{journal}{\emph{arXiv preprint arXiv:1904.09675}} (\bibinfo{year}{2019}).
\newblock


\bibitem[Zhao(2023)]%
        {zhao2023using}
\bibfield{author}{\bibinfo{person}{Wanqun Zhao}.} \bibinfo{year}{2023}\natexlab{}.
\newblock \showarticletitle{Using Science to Support and Develop Employees in the Tech Workforce—An Opportunity for Multidisciplinary Pursuits in Engineering Education}. In \bibinfo{booktitle}{\emph{2023 ASEE Annual Conference \& Exposition}}.
\newblock


\end{thebibliography}

%%%%%%%%%%%%%%%%%%
%figure begins here
%%%%%%%%%%%%%%%%%%
\begin{figure*}
   \centering
    \resizebox{\textwidth}{!}{%
    \includegraphics{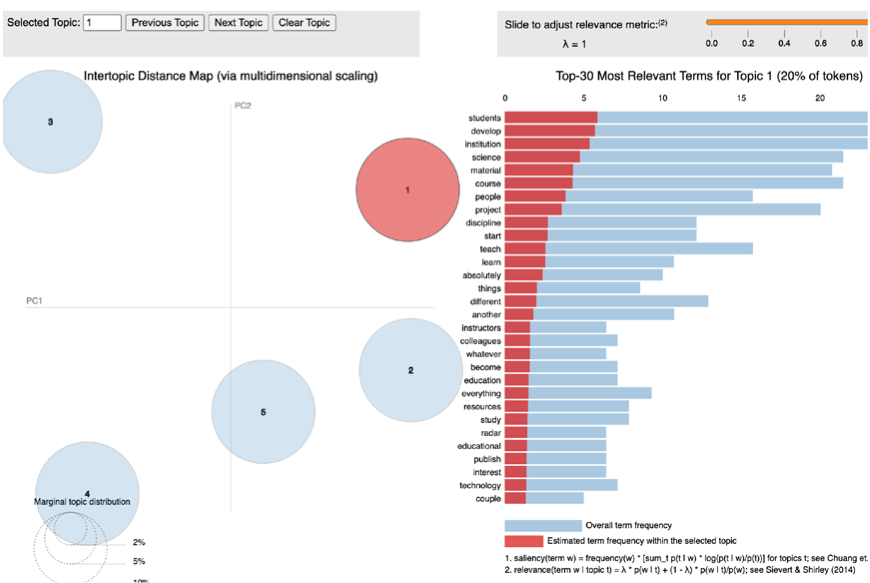}}
    \caption{Topic Modeling using LDA}
\end{figure*}
%%%%%%%%%%%%%%%%%%
%figure ends here
%%%%%%%%%%%%%%%%%%

\appendix
\section{Results from analyzing the same dataset using an LDA Approach.}

Traditional topic modeling using approaches such as Latent Dirichlet Allocation (LDA)  often present the  most representative words for each generated topic. For instance, for Topic 1 words such as "students", "develop", "institution", "science", etc. were found important. Attempting to interpret the underlying thematic meaning of these word lists can be challenging without additional contextual information about how those words were used within the original corpus. In contrast, large language models (LLMs) have demonstrated the capability to synthesize the semantically related words and phrases into more coherent topical representations. This ability of LLMs to generate primitive yet formative contextual information threading together words and phrases of interest and thereby provide researchers with a more insightful starting point for further analysis and interpretation of the latent topics uncovered through the LDA process.

\end{document}